\documentclass{ws-ijgmmp}

\begin{document}

\markboth{Sepp T., Gramann, M.}
{Galaxy structures - groups, clusters and superclusters}

%%%%%%%%%%%%%%%%%%%%% Publisher's Area please ignore %%%%%%%%%%%%%%%
%
\catchline{}{}{}{}{}
%
%%%%%%%%%%%%%%%%%%%%%%%%%%%%%%%%%%%%%%%%%%%%%%%%%%%%%%%%%%%%%%%%%%%%

\title{GALAXY STRUCTURES - GROUPS, CLUSTERS AND SUPERCLUSTERS}

\author{TIIT SEPP}

\address{Department of Cosmology, Tartu Observatory, Observatooriumi 1\\
T\~oravere, Estonia 61602 and Institute of Physics, University of Tartu, T\"ahe 4, Tartu, Estonia 50113\\
\email{ tiit@to.ee } }

\author{MIRT GRAMANN}

\address{Department of Cosmology, Tartu Observatory, Observatooriumi 1\\
T\~oravere, Estonia 61602 \\
mirt@to.ee }

\maketitle

\begin{history}
\received{(30 June 2013)}
\revised{(30 June 2013)}
\end{history}

\begin{abstract}
We provide a brief summary of the history of galaxy structure studies. We also introduce several large-scale redshift surveys and summarize the most commonly used methods to identify the groups and clusters of galaxies. We present several catalogues of galaxy groups.These catalogues can be used to study the galaxy groups in different environments. We also consider the properties of superclusters of galaxies in the nearby Universe and describe the largest system of galaxies observed - the Sloan Great Wall.
\end{abstract}


\section{Introduction}	

Galaxies are the best tracers we have for the study of the structure of the Universe. Although we can consider our Universe homogeneous and isotropic on the largest scales it becomes highly structured once we start to study it in detail. At the first step from largest to smaller the Universe can be described by the cosmic web structure. The cosmic web consists of galaxy-rich areas that contain different galaxy structures - galaxy groups, clusters and superclusters. The second part of the cosmic web contains nearly completely empty voids. This means that matter distribution follows a clearly hierarchical pattern. Analyse of the properties of the different levels of structure  -- galaxy groups, clusters and superclusters, can lead us to better understanding of the evolution in the Universe. 
 
Study of the galaxy structures started on the 18. century, when Charles Messier and F. Wilhelm Herschel independently produced the first catalogues of nebulae. W. Herschel was one of the firsts who suggested that many nebula that we see are external to our own Galaxy. He also recognized several nearby clusters and groups of galaxies. His work was continued by his son, John F.W. Herschel, who already hinted at the existence of the Local Supercluster. In the beginning of the 20. century, it became clear that many nebulae are indeed of extragalactic origin. In 1923 Reynolds made the first reference to the existence of Local Supercluster, with the Virgo cluster as the main body of the system \cite{rey1}. In 1927 Lundmark studied the large-scale distribution of 55 clusters and noted that the most characteristic feature in the their distribution is the clustering tendency \cite{lundmark}. Many nearby galaxy clusters and superclusters were discovered in years 1930-1940. In 1933 F. Zwicky first estimated the mass of the galaxy cluster, thus establishing the need for dark matter\cite{zwicky}.

After the Second World War, the Lick and Palomar sky surveys and the new spectroscopic observations provided the important database for the analysis of the large-scale distribution of galaxies. The search for nearby galaxy structures become more systematic. In 1958 Abell published the catalogue of galaxy clusters \cite{abell}. This publication opened a new era in the investigation of galaxy clusters. Abell's 2712 clusters were selected using the red Palomar survey plates because he realized the advantage of the red band over the blue band for the identification of distant clusters. Abell's subjective selection criteria were also very well chosen. He knew that his cluster sample was incomplete at the low richness end, for this reason he defined a statistical subsample of the richest 1682 clusters. Few years later Zwicky et al. \cite{zwicky2} published also the catalogue of galaxies and clusters, but this catalogue did not exert such a large influence on the study of clusters. The main problem in this catalogue was that the sizes of clusters were distance-dependent and this catalogue could not be used as a statistical homogeneous cluster catalogue.

During that time it was realized that galaxy clusters were not the largest structures in the Universe and thus the study of superclusters began. In 1962 Abell \cite{abell2} published the first list of superclusters (although it contained just 17 superclusters). In 1978 J\~oeveer et al. \cite{joeveer} described Perseus and other eight superclusters and noted that majority of clusters form chains (filaments). Einasto et al. \cite{einasto} noted that the large-scale structure in the Universe resembles cellular structure, with galaxies and galaxy clusters concentrated towards cell walls. Large-scale numerical simulations, where structure develops via gravitational instability, confirmed this picture. A more detailed description on the history of cluster and supercluster studies has been written by Biviano \cite{biviano}, where also an extensive list of references can be found.

\section{Data}

In order to study the large-scale distribution of galaxies, we need to know the distance of a large number of galaxies. There exist many different distance indicators, but most of them are not useful to build large and deep catalogues of galaxies because their uncertainties are too large when applied to objects lying at more than 50 Mpc. Fortunately, the cosmological expansion of the Universe, provides us with a practical alternative. By Hubble's law we know that the redshift of a galaxy will be proportional to its distance, so we can use redshift as a substitute for distance. Galaxies have also the peculiar velocities with respect to the Hubble flow and there a number of methods to recover real galaxy distances from their redshifts. By combining redshifts with angular position data, we can study the 3D distribution of galaxies within a field of the sky. 

Redshifts surveys in the 1980s and the 1990s measured from thousands to tens of thouands galaxy redshifts. The first wide-angle redshift survey, which reached beyond the Local Supercluster, was the CfA Redshift Survey \cite{huchra1}. It started in 1977 and lasted until 1982, measuring 2400 redshifts. The extension of the CFA, the CfA2 survey (1985-1995), on time of completion 18000 redshifts. Now the multi fibre technology allows to measure redshifts of millions of galaxies. The large redshift surveys are usually magnitude limited and this means that in more distant regions we see only brighter galaxies. There are different methods to handle this problem. One possibility is to calculate the corrected luminosities to compensate the loss of not observed fainter galaxies. Advantage of such a method is the maximum use of observational data. However, there are studies, which need to create volume-limited samples. In this case we can fix an absolute magnitude limit for a galaxy sample and reject all less luminous galaxies. 

\begin{figure}[ph]
\centerline{\psfig{file=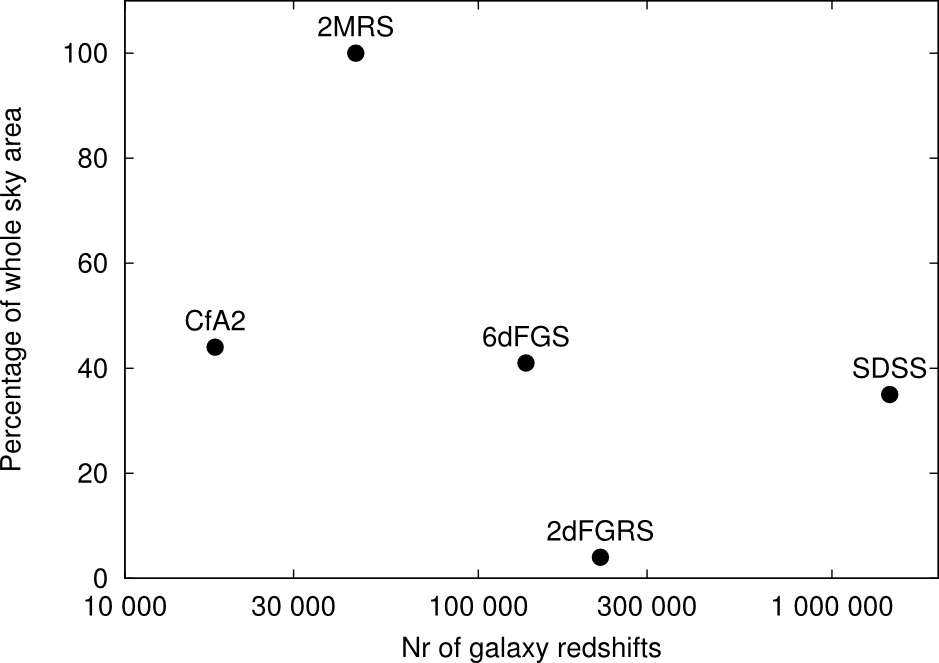, width=4in}}
\vspace*{8pt}
\caption{Different redshift galaxy surveys. \label{comp}}
\end{figure}

Below we summarize briefly the properties of the recent large redshift surveys - 2dFGRS, SDSS, 6dFGS and 2MRS. These are the surveys that cover more than 1000 square degrees in the sky. The number of galaxy redshifts and the survey area for different redshift surveys are given also in Fig.~\ref{comp}.

The 2dF galaxy redshift survey (2dFGRS) measured redshifts for approximately 220000 galaxies using the 2dF multi-object spectrograph on the Anglo-Australian Telescope (AAT) \cite{colless}. It started in 1997 and lasted until 2002. The galaxies with magnitudes brighter than $m_B < 19.45$ were selected from the APM galaxy catalog. This survey covers a region about 1500 square degrees, which is substantially smaller than in the other large redshift surveys as SDSS and 6dFGS. However, the 2dFGRS survey is deeper than 6dFGS and SDSS in its present stage. The mean redshift in the 2dFGRS survey is z = 0.1.

The Sloan Digital sky survey (SDSS) is the largest galaxy redshift survey to date. This survey is carried out using a dedicated 2.5m telescope at Apache Point Observatory in New  Mexico. During its first phase of operations (SDSS-I, 2000-2005), the SDSS imaged more than 8000 square degrees of the sky in five optical bandpasses and obtained spectra of galaxies selected from 5700 square degrees of that imaging. During its second phase (SDSS-II, 2005-2008), the SDSS completed its original goals and measured around 900000 galaxy redshifts in the region about 8000 square degrees. The spectra were obtained for galaxies brighter than $m_r < 17.77$. At present, the SDSS is continuing its third phase (SDSS-III), which began to collect data in 2008 and is planned to continue until 2014. The latest data release (Data Release 9) includes about 1.5 million redshifts in the region about 14500 square degrees \cite{ahn}. However, this huge dataset is not yet homogeneous. To study the large-scale distribution of galaxies, it is recommended to use the main continuous region in the North Galactic Cap, that covers about 7200 square degrees (the Legacy Survey). This region contains about 500000 galaxies. 
 
The 6dF galaxy survey (6dFGS)and 2MASS redshift survey (2MRS) are both near-infrared selected surveys and base on the Two Micron All Sky Survey (2MASS). The 2MASS is the 
photometric survey which mapped almost all of the sky (99.998\%) in the near infra-red 
J, H and K-bands \cite{skrutskie}. The 6dF galaxy redshift survey (6dFGS) measured redshifts for around $\sim 125000$ galaxies in the 2MASS galaxy catalog to a limiting magnitude of $K = 12.75$ \cite{jones}. The redshift were obtained using the 6dF multi-object spectrograph on the UK Schmidt Telescope in Australia. The 6dFGS survey covers a region about 17000 square degrees on the Southern hemisphere. 

The 2MASS redshift survey (2MRS) ultimately aims to determine the redshifts of all galaxies in the 2MASS database to a magnitude of $K = 12.2$ (about 100000 galaxies) \cite{huchra2}. The second phase of 2MRS is now complete, providing an all-sky redshift survey of 45000 galaxies to a limiting magnitude of $K = 11.75$. This survey is the densest sampled all-sky redshift survey to date and its selection in the near infra-red reduces the impact of the zone of avoidance (where the plane of our own Galaxy obscures extragalactic objects). 2MRS provides complementary redshift information to deeper surveys like SDSS and the 2dFRGS which cover smaller fractions of the sky.

\section{Galaxy groups and clusters}

Observations of the local Universe show that almost half of the galaxies are located in groups. During the evolution galaxy groups fall into the rich clusters of galaxies, which can contain several smaller groups within. Rich galaxy clusters are largest gravitationally bound galaxy systems and their contain about 10 per cent of the galaxies. Therefore, the groups and clusters are natural environment for galaxies, and their study is important for understanding the evolution of galaxies and the underlying matter distribution. Below we give a brief overview about the different group catalogues obtained using different redshift surveys. Selection of group catalogues is listed in Table 1. The number $N_{gr} (n \geq 3)$ is the number of groups which contain at least three galaxies.

Abell's cluster catalog \cite{abell} was created by visual inspection of the Palomar photographic plates. With the help of large redshift surveys, we can identify groups in three dimensional space. The most frequently used method to date is the friends-of-friends method (FoF, sometimes called the percolation method). With the FoF method, galaxies are linked into systems, using a certain linking length (or neighbourhood radius). 

\begin{table}[ht]
\tbl{Selection of group catalogues}
{\begin{tabular}{@{}cccccc@{}} \toprule
Main author \& year & reference & $N_{gr} (n \geq 3)$ &  Data & Method
\\ \colrule
\\
Geller \& Hurcha, 1983 & \cite{geller} & 176 & CfA & FoF \\
Eke, 2004 & \cite{eke} & 12566 &  2dFGRS & FoF \\
Crook, 2007 & \cite{crook} & 1538 &  2MRS & FoF\\
Berlind, 2006 & \cite{berlind} & 4107 & SDSS & FoF \\
Yang, 2007 & \cite{yang} & 9949 &  SDSS & mass-halo \\
Tempel, 2012 & \cite{tempel} & 30515 &  SDSS & FoF \\
Sepp, 2013 & \cite{sepp} & 18507 &  SDSS & density field \\
\\ \botrule
\end{tabular}}
\label{table}
\end{table}

In 1983 Geller and Huchra \cite{geller} used FoF method to identify groups in the CfA redshift survey. They identified 176 nearby galaxy groups which have three or more members. Eke et al. (2004) \cite{eke} studied the galaxy groups in the 2dFGRS survey. The groups were identified by means of FoF method which was tested by using the cosmological N-body simulations. About 55 per cent of the galaxies considered were found to be in groups containing at least two members. Of these, $\sim 12500$, contain at least three galaxies. Crook et al. (2007) \cite{crook} used FoF method to identify groups in the 2MRS survey. They presented two catalogues of groups identified with slightly different linking lengths. The first catalogue was created by maximizing the number of groups containing three or more members. The second catalogue was created with smaller linking length to identify the largest nearby clusters individually. Berlind et al. (2006)\cite{berlind} identified galaxy groups in volume limited samples of the SDSS redshift survey. They optimized the FoF method by using a set of mock catalogs created by populating halos of N-body simulations with galaxies. Tempel et al. (2012) \cite{tempel} constructed a group catalogue for the SDSS DR8 sample. They used a modified FoF method with variable linking lengths to minimize the selection effects in the magnitude-limited sample. Their final sample contains 30515 galaxy groups with three or more galaxies. About 46 per cent of all galaxies were found to be in groups containing at least two members. 

Two other commonly used methods to identify the groups of galaxies in redshift surveys are halo-based method and the density-field method. The halo-based method was used by Yang et al. (2007) \cite{yang} to select galaxy groups from the SDSS redshift survey. In the first step, this method uses the traditional FoF algorithm to find potential group centers and then estimates the properties of the halos associated with each groups. This halo information is used to determine the group membership in redshift space. The second method to identify the groups, the density-field method, bases on the smoothed density field of galaxies. This method was used by Sepp et al. 2013 in the SDSS redshift survey \cite{sepp}. Here the groups of galaxies are identified at the maxima of the smoothed density field of galaxies and then the FoF method is applied to determine the group membership around density maxima. 

\section{Galaxy superclusters}

Galaxy superclusters consist of galaxies, groups and clusters of galaxies connected by filaments. They form the basic building blocks of the structure in the Universe. Most of the rich clusters are located in the superclusters and many very luminous clusters are in the cores of superclusters. Studies of the galaxies and galaxy systems in different superclusters are important for understanding the evolution of galaxy groups and the large-scale structure in the Universe. They are still in the beginning of virialization and in the distant future they will evolve into compact isolated systems as postulated by Araya-Melo et al. \cite{araya}. From topological studies \cite{park} we know that there exist small deviations from linear gravitational evolution and these effects can be studied using galaxy superclusters.

Early supercluster catalogues have been compiled using data on clusters of galaxies. The superclusters of Abell clusters were identified by Einasto et al. (2001) (E01) \cite{einasto1}. With the advent of large redshift surveys we can study the large-scale structure of galaxies in much more detail. Liivam\"agi et al. (2012) (L12 - from ) studied the the large-scale distribution of galaxies in the SDSS redshift survey and constructed a set of supercluster catalogues L12 \cite{liivamagi}. To delineate the superclusters, they calculated the large-scale luminosity density field and found regions with densities over a selected threshold. A similar method to identify the superclusters in the SDSS survey was used also by Costa-Duarte et al. (2011) \cite{costa} and Luparello et al. (2011)\cite{luparello}. Alternatively, the superclusters can be find also with the FOF method. This method was used by Basilakos \cite{basilakos} to compile superclusters of galaxies from the SDSS survey. 

The Sloan redshift survey covers many nearby rich superclusters as the Hercules supercluster and the Corona Borealis supercluster. The richest and largest system of galaxies observed in the nearby Universe is the Sloan Great Wall (SGW) \cite{gott}, \cite{einasto2}. The mean distance to this large system of galaxies is about 230 $h^{-1}$ Mpc. Fig. 2 shows the distribution of galaxies in the Sloan Great Wall. The distribution is shown in cartesian coordinates x,y,z (defined as in L12). The numbers are ID numbers of superclusters in the L12 catalogue for the SDSS DR8 survey. We can see the size of the Sloan Great Wall, its linear size is about 300 $h^{-1}$ Mpc. The SGW consists of several superclusters of galaxies; the richest of them is the supercluster SCL 27 (SCL 126 in E01), which contain many rich clusters of galaxies. Another very interesting supercluster in this system is SCL19 (SCL 111 in E01). The superclusters in the SGW differ in morphology and galaxy content, which suggests that their formation and evolution has been different.

\begin{figure}[ph]
\centerline{\psfig{file=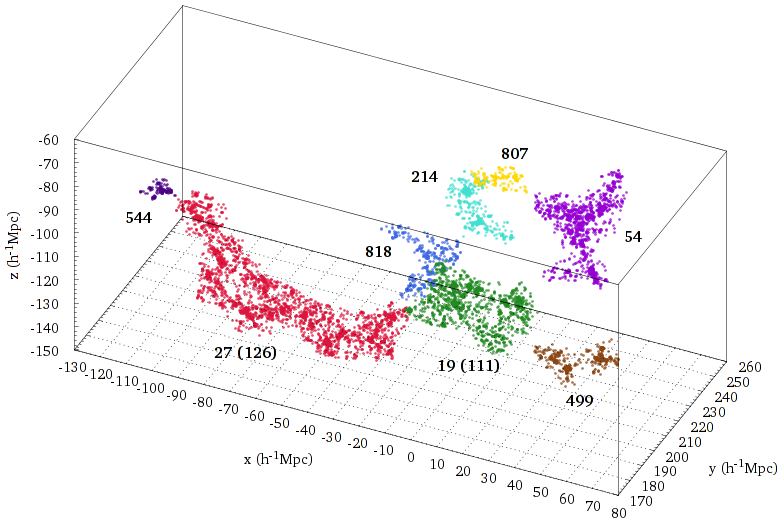, width=4in}}
\vspace*{8pt}
\caption{Distribution of galaxies in different superclusters in the Sloan Great Wall.  credit L. J. Liivam\"agi . \label{sss} }
\end{figure}

\section*{Acknowledgements}

We would like to thank the organizers of 49th Winter School of Theoretical Physics
Cosmology and non-equilibrium statistical mechanics. Also we would like to thank L. J. Liivam\"agi for stimulating discussions about superclusters and providing us with Fig ~\ref{sss}.

%\appendix

%\section{Appendices}

%\section*{References}

\end{document}